\documentclass[11pt,twoside,english]{article}
\usepackage[T1]{fontenc}
\usepackage[latin9]{inputenc}
\usepackage[a4paper]{geometry}
\usepackage{fancyhdr}
\usepackage{babel}
\usepackage{lmodern}
\usepackage[babel=true]{microtype}
\usepackage{mathrsfs}
\usepackage{amsmath}
\usepackage{amsthm}
\usepackage{amssymb}
\usepackage{graphicx}
\usepackage[unicode=true,pdfusetitle,
 bookmarks=true,bookmarksnumbered=false,bookmarksopen=false,
 breaklinks=false,pdfborder={0 0 0},pdfborderstyle={},backref=false,colorlinks=false]
 {hyperref}
\hypersetup{
 pdfpagemode=UseNone,pdfstartview=FitH,pdfborderstyle={/S/U/W 1}}

\makeatletter
\theoremstyle{plain}
\newtheorem*{thm*}{\protect\theoremname}

\usepackage[small,bf]{caption2}
\usepackage{graphicx}
\usepackage{wrapfig}
\usepackage[scaled]{helvet}
\usepackage[usenames,dvipsnames]{color}%
\usepackage{floatrow}
\usepackage{overpic}
\usepackage[caption=false]{subfig}

\setcaptionmargin{0cm}

\renewenvironment{thebibliography}[1]{   
\begin{oldthebibliography}{#1}     
\setlength{\itemsep}{1mm}     
\setlength{\parskip}{0em} } {   
\end{oldthebibliography} }

\makeatother

\providecommand{\theoremname}{Theorem}

\begin{document}
\fancyhead[OL]{} 
\fancyhead[OR]{} 
\fancyhead[ER]{} 
\fancyhead[EL]{} 
\fancyfoot[EL]{}\fancyfoot[EC]{\thepage} 
\fancyfoot[OR]{}\fancyfoot[OC]{\thepage} 
\renewcommand{\headrulewidth}{0.0pt}% 
\title{On the Holway-Weiss Debate: Convergence of the Grad-Moment-Expansion
in Kinetic Gas Theory}
\author{
Zhenning Cai\\
Department of Mathematics,\\
National University of Singapore,\\
10 Lower Kent Ridge Rd, Singapore 119076, Singapore,\\
\href{mailto:matcz@nus.edu.sg}{matcz@nus.edu.sg}\\
\\
Manuel Torrilhon\\
Center for Computational Engineering Science,\\
RWTH Aachen University,\\
Schinkelstr. 2, 52062 Aachen, Germany,\\
\href{mailto:torrilhon@rwth-aachen.de}{torrilhon@rwth-aachen.de}
}
\date{(2019)}
\maketitle
\begin{abstract}
Moment expansions are used as model reduction technique in kinetic
gas theory to approximate the Boltzmann equation. Rarefied gas models
based on so-called moment equations became increasingly popular recently.
However, in a seminal paper by Holway {[}Phys.~Fluids \textbf{7}/6,
(1965){]} a fundamental restriction on the existence of the expansion
was used to explain sub-shock behavior of shock profile solutions
obtained by moment equations. Later, Weiss {[}Phys.\,Fluids \textbf{8}/6,
(1996){]} argued that this restriction does not exist. We will revisit
and discuss their findings and explain that both arguments have a
correct and incorrect part. While a general convergence restriction
for moment expansions does exist, it cannot be attributed to sub-shock
solutions. We will also discuss the implications of the restriction
and give some numerical evidence for our considerations.
\end{abstract}

\section{Introduction}

\fancyhead[OL]{Z.~Cai and M.~Torrilhon} 
\fancyhead[OR]{On the Holway-Weiss-Debate} 
\fancyhead[ER]{Z.~Cai and M.~Torrilhon} 
\fancyhead[EL]{On the Holway-Weiss-Debate} 
\fancyfoot[EL]{}\fancyfoot[EC]{\thepage} 
\fancyfoot[OR]{}\fancyfoot[OC]{\thepage} 
\renewcommand{\headrulewidth}{0.4pt}% 

Rarefied gas dynamics is encountered in a number of modern technological
fields such as high-altitude spacecrafts and microelectromechanical
systems, and the modeling of rarefied gases has been of interest for
more than one century. It is generally agreed that the Boltzmann equation,
the fundamental equation in gas kinetic theory \cite{Cercignani1},
provides an accurate description for rarefied gases in most applications.
However, simulations using the Boltzmann equation require to solve
a six-dimensional distribution function, which is computationally
expensive and often unaffordable. Therefore, researchers have been
trying to derive cheaper models from the Boltzmann equation by model
reduction. A classical approach is the moment method, which was first
introduced to gas kinetic theory by H.~Grad in \cite{Grad1949}.
Grad suggested to expand the distribution function in velocity
space using orthogonal polynomials with the local Maxwellian as the
weight function, and derived a 13-moment model by a truncation of
such expansion. 

In the past three decades many successful results in the modeling
of rarefied gases could be obtained based on Grad's approximation.
Starting with the close relation to phenomenological extended thermodynamics
(ET), Grad's moment equations were used to compute sound waves, light
scattering and shock profiles in the context of ET, see \cite{Muller1998}.
The regularized theory of \cite{Struchtrup2003} allows to formulate
consistent boundary conditions \cite{GuEmerson,torrilhon2008boundary}
based on Grad's distribution. This leads to the successful application
of Grad's moment equations to a variety of boundary value problems
\cite{torrilhon2016modeling}. Recently, Grad's distribution has
also been used with very high number of moments such that moment
equations become a numerical technique to solve the Boltzmann equation
efficiently and accurately \cite{Cai2010,TorrilhonSarna2017}. 

In the early days of Grad's theory it was far from obvious that the
theory was actually useful. In \cite{Grad1949} H.~Grad reports
about artefacts in the shock profile computation obtained with his
13-moment-equations and concludes that the new equations do not show significant
improvement over classical theories. The seminal paper of Holway \cite{Holway1964}
links the shock artefacts to a mathematical statement about the general
convergence of moment expansions and concludes that moment equations
cannot be used beyond a certain shock strength given by a critical
Mach number. However, W.~Weiss computed smooth shock profiles beyond
the critical Mach number in \cite{Weiss1995} and subsequently published
a paper \cite{Weiss1996} in which he disproves the statement
of Holway about the usefulness of moment equations in shock waves. 

In recent years computations of the authors of this paper made it
clear that the Holway-Weiss-debate deserves to be revisited, because
the current representation in the literature is highly misleading.
After introducing the foundations of Grad's expansion in Sec.~2 we
will carefully reframe the argument of Holway in Sec.\,3 and discuss
in detail in Sec.\,3.2 what is right or wrong in both Holway's statement
and Weiss' rebuttal. In Sec.~4 we discuss the consequences of the
argument for nowadays moment models and give some numerical examples.

\section{Basic Functional Analysis of the Grad Expansion}

Given the density $\rho$, velocity vector $v_{i}$ and temperature
$\theta$ (in energy density units) of the gas the simplest approximation
for the underlying distribution function is to assume equilibrium.
The Maxwell distribution for the particle velocities $c_{i}$
\begin{align}
f_{\text{eq}}(\mathbf{c}) & =\frac{\rho/m}{(2\pi\theta)^{3/2}}\exp\left(-\frac{(c_{i}-v_{i})^{2}}{2\theta}\right)
\end{align}
then gives a distribution that fits the density, velocity and temperature.
Here $m$ denotes the mass of a single gas molecule. For more general
situations H.~Grad proposed in 1949 to use an expansion which in the simplest
form reads 
\begin{align}
f_{G}^{(M)}(\mathbf{c})=\sum_{\left|\alpha\right|=0}^{M}\lambda_{\alpha}\mathbf{c}^{\alpha}f_{\text{eq}}(\mathbf{c})\label{eq:GradExpansion}
\end{align}
and became known as the Grad distribution. The indices $\alpha$ represent
multi-indices such that $\mathbf{c}^{\alpha}$ are multi-variate monomials
up to degree $M$ and the coefficients $\lambda_{\alpha}$ parametrize
the distribution. They are computed from the consistency requirement
\begin{align}
u_{\alpha} & =\int_{\mathbb{R}^{d}}\mathbf{c}^{\alpha}f_{G}^{(M)}(\mathbf{c})d\mathbf{c}\quad\qquad\left|\alpha\right|=0,1,\ldots, M
\end{align}
with given moments $u_{\alpha}$. Using this technique it is possible
to reconstruct an approximation to the velocity distribution whenever
a set of moments is known. For instance this approximative distribution
$f_{G}^{(M)}$ can be used to close the transfer equations of the
$u_{\alpha}$ obtained from the Boltzmann equation which gives Grad's
moment equations.

\subsection{Best Approximation}

When using orthogonal polynomials instead of monomials the mathematical
analysis of the Grad expansion becomes easier. In fact, the Grad distribution
represents a best approximation in an appropriate subspace. We recall
the following basic theorem, a variant of which could be found, e.g.,
in \cite{ApproxTheory}.
\begin{thm*}[Best Approximation]
Consider the weighted space $V_{\omega}:=L^{2}(\mathbb{R}^{d};\mathbb{R},\omega\,d\mathbf{c})$
with scalar product 
\begin{align}
\left\langle f,g\right\rangle _{\omega} & =\int_{\mathbb{R}^{d}}f(\mathbf{c})g(\mathbf{c})\omega(\mathbf{c})d\mathbf{c}
\end{align}
for functions on $\mathbb{R}^{d}$, and a set of $\omega$-orthonormal
functions $U_{M}=\{\Psi_{\alpha}\}_{\left|\alpha\right|=0,1,...M}\subset V_{\omega}$.
For any function $f$ with $\left\Vert f\right\Vert _{\omega}<\infty$
we find the unique best approximation
\begin{align}
f^{(M)}={\displaystyle \sum}_{\left|\alpha\right|=0}^{M}\left\langle \Psi_{\alpha},f\right\rangle _{\omega}\Psi_{\alpha} & =\underset{f^{\star}\in\text{span}\thinspace U_{M}}{\text{argmin}}\left\Vert f-f^{\star}\right\Vert _{\omega}
\end{align}
in the subspace spanned by $U_{M}$.
\end{thm*}
In the Grad expansion the coefficients of $f_{G}^{(M)}$ are moments
of $f$ which requires the choice $\Psi_{\alpha}(\mathbf{c})=\psi_{\alpha}(\mathbf{c})\omega(\mathbf{c})^{-1}$ with
polynomial functions $\psi_{\alpha}(\mathbf{c})$ in order to find
\begin{align}
w_{\alpha} & =\left\langle \Psi_{\alpha},f\right\rangle _{\omega}=\int_{\mathbb{R}^{d}}\psi_{\alpha}(\mathbf{c})f(\mathbf{c})d\mathbf{c}\quad\Rightarrow\quad f_{G}^{(M)}(\mathbf{c})=\sum_{\left|\alpha\right|=0}^{M}w_{\alpha}\psi_{\alpha}(\mathbf{c})\omega(\mathbf{c})^{-1}
\end{align}
for the expansion. The natural choices for the inverse $\omega(\mathbf{c})^{-1}$ is
a local Maxwellian $f_{\text{eq}}$ with density $\rho$, velocity
$v_i$, and temperature $\theta$ obtained from the respective local
distribution $f$. For more insight into how Grad's expansion is used
in modern gas modeling we refer to the textbook \cite{StruchBook}.
The polynomial functions $\psi_{\alpha}(\mathbf{c})$ satisfy the orthogonality
condition
\begin{align}
\left\langle \Psi_{\alpha},\Psi_{\beta}\right\rangle _{\omega} & =\int_{\mathbb{R}^{d}}\psi_{\alpha}(\mathbf{c})\psi_{\beta}(\mathbf{c})f_{\text{eq}}(\mathbf{c})d\mathbf{c}=\begin{cases}
1\qquad & \alpha=\beta\\
0\qquad & \text{else}
\end{cases}
\end{align}
such that they are nothing but Hermite polynomials with the Gaussian
of the local equilibrium distribution as weight. The weighted functions
$\Psi_{\alpha}=\psi_{\alpha}\omega^{-1}=\psi_{\alpha}f_{\text{eq}}$
are sometimes called \emph{Hermite functions}. 

The following theorem is the reason for a fundamental condition for
the existence of a Grad expansion.
\begin{thm*}[Completeness]
The infinite set of Hermite functions $\{\Psi_{\alpha}\}_{\alpha\in\mathbb{N}^{d}}$
form a complete orthogonal basis of the weighted space $V_{\omega}$
with $\omega=f_{\text{eq}}^{-1}$. 
\end{thm*}
This statement means any function $f\in V_{\omega}$ with $\omega=f_{\text{eq}}^{-1}$ can
be represented through a Grad expansion and vice versa. Hence, if
$f$ is to be represented as an infinite Grad expansion, necessarily
$\left\Vert f\right\Vert _{\omega}<\infty$ must hold.
\begin{figure}[!t]
\subfloat[$\mathit{Ma} = 3$]   {\includegraphics[height=.23\textwidth]{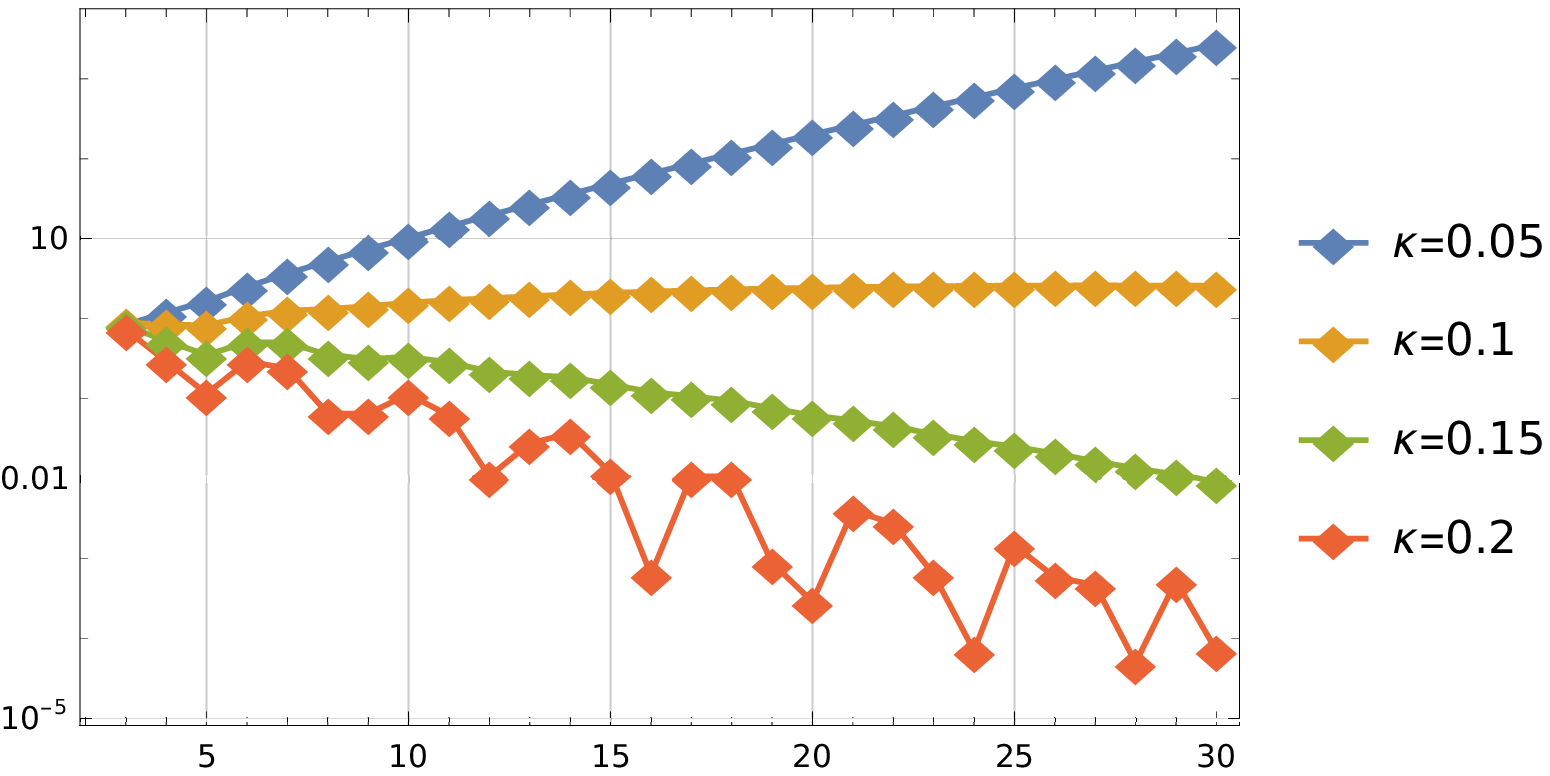}} \hspace{15pt} \subfloat[$\kappa = 0.1$]   {\includegraphics[height=.23\textwidth]{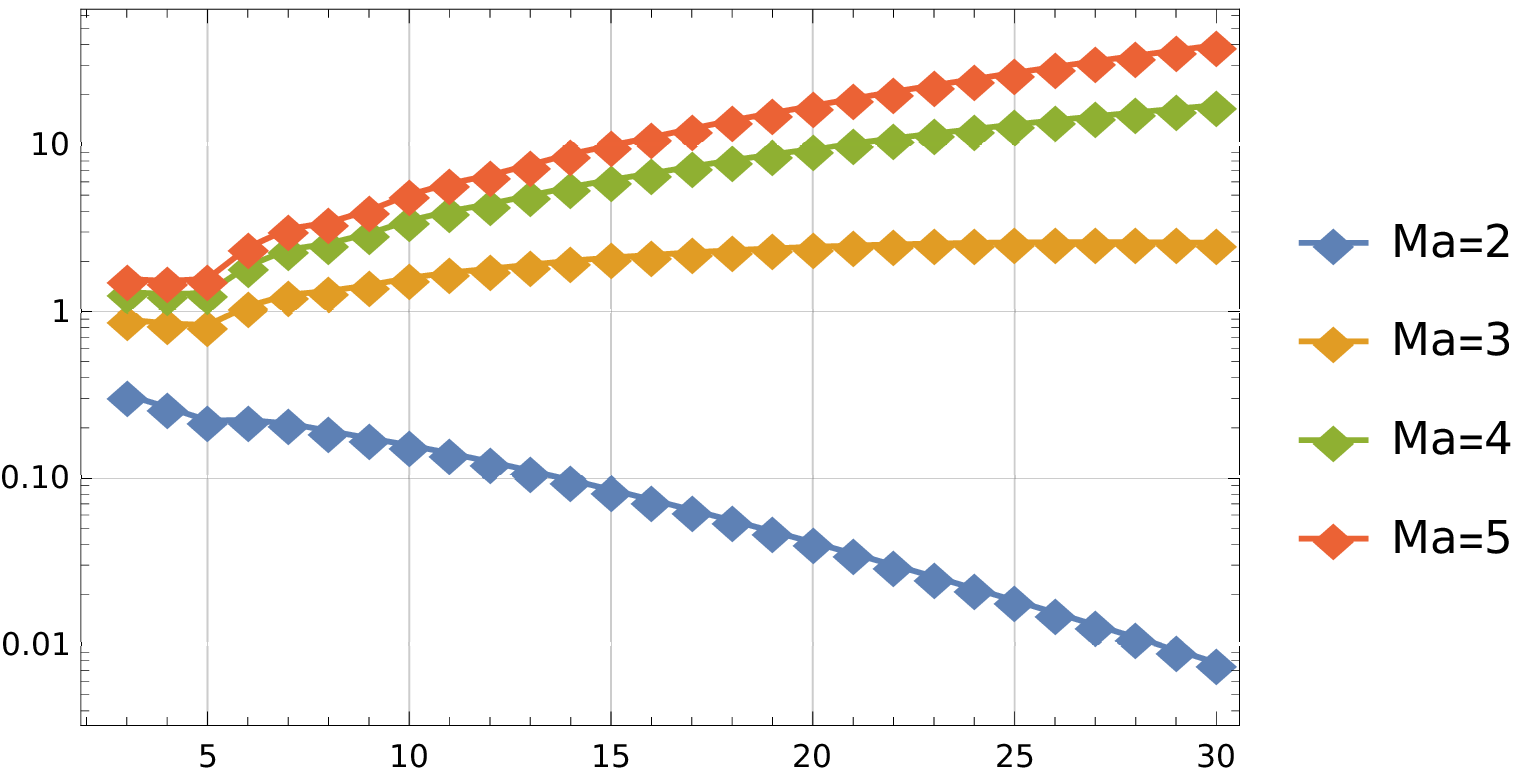}} 
\caption{Absolute value of the expansion coefficients $|w_{\alpha}|$ for \textbf{(a)} fixed 
value of $\mathit{Ma}$ and different $\kappa$, and \textbf{(b)} fixed value of 
$\kappa$ and different $\mathit{Ma}$. The horizontal axis shows the value
of the coefficient index $\alpha$, while the vertical axis is the
log scale of $|w_{\alpha}|$.}
\label{fig:coef} 
\end{figure}

\subsection{Restrictions on the Distribution}

The condition $\left\Vert f\right\Vert _{\omega}<\infty$ leads to
\begin{align}
\int_{\mathbb{R}^{d}}\left(ff_{\text{eq}}^{-1/2}\right)^{2}d\mathbf{c} & <\infty,
\end{align}
hence the decay rate of $f$ must be faster than that of $f_{\text{eq}}^{1/2}$.
In other words, if we let $\theta^{(\text{tail})}$ be the tail temperature
of $f$ such that 
\begin{align}
f(c)\leq R\exp\left(-\frac{\mathbf{c}^{2}}{2\mathcal{\theta}^{(\text{tail})}}\right) & \qquad\text{for }\left\Vert \mathbf{c}\right\Vert > r
\end{align}
with some constants $R$ and $r$, then for $\theta$ being the actual temperature
of $f_{\text{eq}}$ implied by $f$,
\begin{align}
\theta^{(\text{tail})} & <2\theta\label{eq:DecayRestriction}
\end{align}
must hold. In physical terms this means that the fast particles in
the tail of the distribution should not be hotter than twice the average
of all particles in the distribution.
\begin{figure}[!t]
\subfloat[$\mathit{Ma} = 3$, $\kappa = 0.15$]   {\includegraphics[height=.23\textwidth]{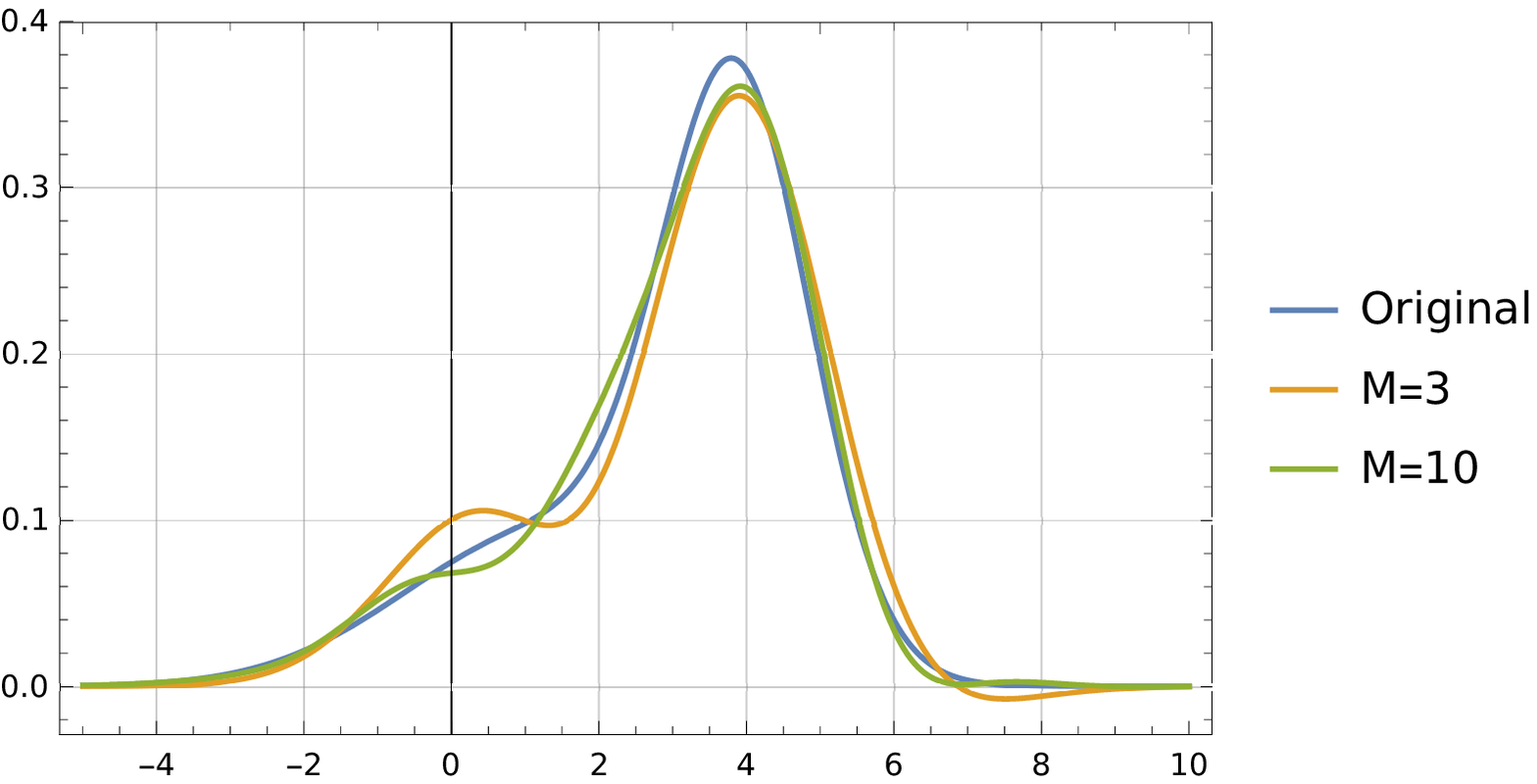}} \hspace{15pt} \subfloat[$\mathit{Ma} = 4$, $\kappa = 0.1$]   {\includegraphics[height=.23\textwidth]{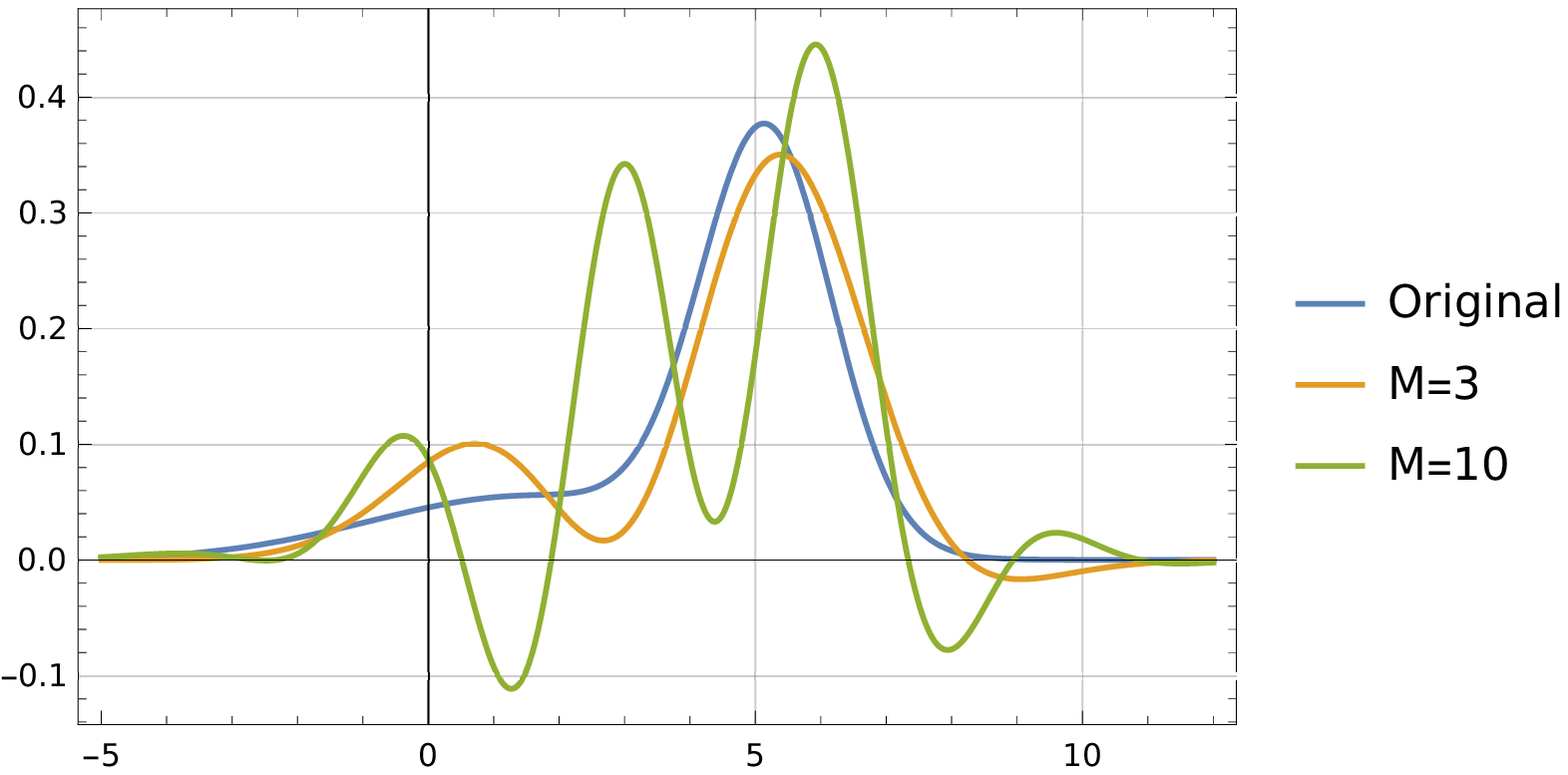}} \caption{1D marginal distribution function for the superposition of two Maxwellians
and the truncated Grad's expansion. Figure \textbf{(a)} shows a convergent case, while in \textbf{(b)} convergence fails.}
\label{fig:dis} 
\end{figure}

\subsection{Convergence Failure\label{sec:failure}}

However, (\ref{eq:DecayRestriction}) does not always hold. One typical example
of distributions that may violate (\ref{eq:DecayRestriction}) is the
superposition of two Gaussians: 
\begin{equation}
f(\mathbf{c})=\frac{\rho_{1}/m}{(2\pi\theta_{1})^{3/2}}\exp\left(-\frac{|\mathbf{c}-\mathbf{v}_{1}|^{2}}{2\theta_{1}}\right)+\frac{\rho_{2}/m}{(2\pi\theta_{2})^{3/2}}\exp\left(-\frac{|\mathbf{c}-\mathbf{v}_{2}|^{2}}{2\theta_{2}}\right).\label{eq:Bimodal}
\end{equation}
The velocity and temperature of such a distribution function can be
calculated directly: 
\begin{equation}
\mathbf{v}=\frac{\rho_{1}\mathbf{v}_{1}+\rho_{2}\mathbf{v}_{2}}{\rho_{1}+\rho_{2}},\qquad\theta=\frac{\rho_{1}\theta_{1}+\rho_{2}\theta_{2}}{\rho_{1}+\rho_{2}}+\frac{\rho_{1}\rho_{2}|\mathbf{v}_{1}-\mathbf{v}_{2}|^{2}}{3(\rho_{1}+\rho_{2})^{2}}.\label{eq:VelTemp}
\end{equation}
Suppose $\theta_{1}<\theta_{2}$. Then the tail of this distribution
function is governed by Maxwellian with temperature $\theta_{2}$.
However, by (\ref{eq:VelTemp}), we can see that by decreasing $\rho_{2}$,
the value of $\theta$ can be set to be arbitrarily close to $\theta_{1}$.
This means that when $\theta_{2}>2\theta_{1}$, for any given $\mathbf{v}_{1}$
and $\mathbf{v}_{2}$, we can always choose $\rho_{2}\ll\rho_{1}$
such that (\ref{eq:DecayRestriction}) does not hold, leading to the
divergence of the Grad expansion.

Such an example is of interest since it comes from the Mott-Smith
bimodal theory for the steady shock structure problem \cite{MottSmith}.
In such a theory, one considers a plane shock wave with Mach number
$\mathit{Ma}$, and it is assumed that the distribution function takes
the form (\ref{eq:Bimodal}) everywhere, with the following dimensionless
parameters: 
\begin{equation}
\begin{gathered}\rho_{1}=1-\kappa,\quad\mathbf{v}_{1}=\left(\sqrt{5/3}\mathit{Ma},0,0\right)^{T},\quad\theta_{1}=1,\\
\rho_{2}=\kappa\frac{4\mathit{Ma}^{2}}{\mathit{Ma}^{2}+3},\quad\mathbf{v}_{2}=\left(\sqrt{\frac{5}{3}}\frac{\mathit{Ma}^{2}+3}{4\mathit{Ma}},0,0\right)^{T},\quad\theta_{2}=\frac{(5\mathit{Ma}^{2}-1)(\mathit{Ma}^{2}+3)}{16\mathit{Ma}^{2}},
\end{gathered}
\label{eq:ShockSetting}
\end{equation}
where $\kappa\in[0,1]$ varies with spatial location. To illustrate
the divergence, we present in Fig.~\ref{fig:coef} the magnitude
of the expansion coefficients for different $\kappa$ and $\mathit{Ma}$,
from which one can observe that the coefficient $|w_{\alpha}|$ does
increase when $\mathit{Ma}$ is large and $\kappa$ is small. In 
Fig.~\ref{fig:dis}, we provide the comparison between the distribution
function and the truncated Grad's expansion. For $\mathit{Ma}=3$
and $\kappa=0.15$, Grad's expansion converges, and the figure shows
that the truncation at $M=10$ gives better approximation than $M=3$.
For $\mathit{Ma}=4$ and $\kappa=0.1$, Grad's expansion diverges.
Although the truncation at $M=3$ still seems to be approximating
the original distribution function, the result given by $M=10$ clearly
shows the failure of convergence.

\section{A General Variant of Holway's Argument}

\begin{figure}
\includegraphics[width=8cm]{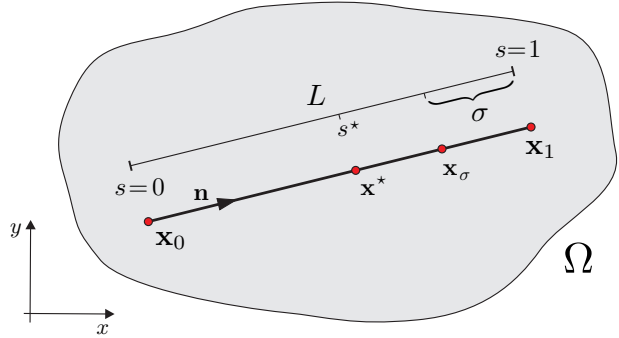}
\caption{Generalized setup of the considerations of Holway. Particles travel
in the direction $\mathbf{n}$ from $\mathbf{x}_{0}$ to $\mathbf{x}_{1}$.
This means that properties like the decay rate of the distribution
at position $\mathbf{x}^{\star}$ influence the distribution at position
$\mathbf{x}_{1}$.}
\label{fig:setup}
\end{figure}

We will generalize the original argument of Holway to the case of
a multi-dimensional domain $\Omega$ in which we consider two arbitrary
points $\mathbf{x}_{0},\mathbf{x}_{1}\in\Omega$ connected by a straight
line of length $L$ and direction unit vector $\mathbf{n}$ as shown
in Fig.~\ref{fig:setup}. While Holway constructed his statement
for the full Boltzmann collision operator, assuming some estimates
for the gain part, we will directly use the BGK approximation to simplify
the presentation. The extension to the Boltzmann operator is of purely
technical nature and is not relevant for our discussion. 

\subsection{Derivation\label{subsec:Derivation}}

The steady BGK\,equation is written with constant collision frequency
for a particle velocity $\mathbf{c}=c\,\mathbf{n}$ pointing from
$\mathbf{x}_{0}$ to $\mathbf{x}_{1}$ 
\begin{align}
c\,\mathbf{n}\cdot\nabla f\left(\mathbf{x},c\,\mathbf{n}\right) & =-\nu\,f\left(\mathbf{x},c\,\mathbf{n}\right)+\nu\,f_{\text{eq}}\left(\mathbf{x},c\,\mathbf{n}\right)
\end{align}
which can be transformed into
\begin{align}
\mathbf{n}\cdot\nabla\left(f\left(\mathbf{x},c\,\mathbf{n}\right)\exp\left(\frac{\nu}{c}\left\Vert \mathbf{x}-\mathbf{x}_{0}\right\Vert \right)\right) & =\frac{\nu}{c}\exp\left(\frac{\nu}{c}\left\Vert \mathbf{x}-\mathbf{x}_{0}\right\Vert \right)f_{\text{eq}}\left(\mathbf{x},c\,\mathbf{n}\right)
\end{align}
after multiplying with $\frac{1}{c}\exp\left(\frac{\nu}{c}\left\Vert \mathbf{x}-\mathbf{x}_{0}\right\Vert \right)$.
If we replace $\mathbf{x}$ by the parametrization of the line $\mathbf{x}(s)=\mathbf{x}_{0}+s\left\Vert \mathbf{x}_{1}-\mathbf{x}_{0}\right\Vert \mathbf{n}$,
we can integrate from $\mathbf{x}(0)=\mathbf{x}_{0}$ to $\mathbf{x}(1)=\mathbf{x}_{1}$,
which gives

\begin{align}
\int_{0}^{1}\mathbf{n}\cdot\nabla\left(f\left(\mathbf{x}(s),c\,\mathbf{n}\right)\exp\left(\frac{\nu}{c}\left\Vert \mathbf{x}(s)-\mathbf{x}_{0}\right\Vert \right)\right) & \left\Vert \mathbf{x}'(s)\right\Vert ds=\\
 & \frac{\nu}{c}\int_{0}^{1}\exp\left(\frac{\nu}{c}\left\Vert \mathbf{x}(s)-\mathbf{x}_{0}\right\Vert \right)f_{\text{eq}}\left(\mathbf{x}(s),c\,\mathbf{n}\right)\left\Vert \mathbf{x}'(s)\right\Vert ds.\nonumber 
\end{align}
Replacing $\left\Vert \mathbf{x}'(s)\right\Vert =\left\Vert \mathbf{x}_{1}-\mathbf{x}_{0}\right\Vert =L$
and $\mathbf{n}\cdot\nabla\rightarrow\partial_{s}$ we can compute
the integral on the left hand side explicitly and find 
\begin{align}
f\left(\mathbf{x}_{1},c\,\mathbf{n}\right)=f\left(\mathbf{x}_{0},c\,\mathbf{n}\right)\exp\left(-\frac{\nu L}{c}\right)+\frac{\nu L}{c}\int_{0}^{1}\exp\left(-\frac{\nu L}{c}(1-s)\right)f_{\text{eq}}\left(\mathbf{x}(s),c\,\mathbf{n}\right)ds
\end{align}
after multiplication with $\exp\left(-\frac{\nu L}{c}\right)$. Note
that all terms in this equation are positive.
Holway then integrates from $\mathbf{x}_{0}$ to $\mathbf{x_{\sigma}}=\mathbf{x}(1-\sigma)$
with $0<\sigma<1$ and writes 
\begin{align}
f\left(\mathbf{x}_{1},c\,\mathbf{n}\right) & \geq\frac{\nu L}{c}\int_{0}^{1-\sigma}\exp\left(-\frac{\nu L}{c}(1-s)\right)f_{\text{eq}}\left(\mathbf{x}(s),c\,\mathbf{n}\right)ds
\end{align}
to be used as lower bound of $f$ at \textbf{$\mathbf{x}_{1}$}. Using
the mean value theorem with an $s^{\star}\in(0,1-\sigma)$, and $\mathbf{x}(s^{\star})=\mathbf{x}^{\star}$
we find
\begin{align}
f\left(\mathbf{x}_{1},c\,\mathbf{n}\right) & \geq f_{\text{eq}}\left(\mathbf{x}^{\star},c\,\mathbf{n}\right)\frac{\nu L}{c}\int_{0}^{1-\sigma}\exp\left(\frac{\nu L}{c}(s-1)\right)ds\\
 & =f_{\text{eq}}\left(\mathbf{x}^{\star},c\,\mathbf{n}\right)\left(\exp\left(-\frac{\nu L}{c}\sigma\right)-\exp\left(-\frac{\nu L}{c}\right)\right)
\end{align}
hence obtain
\begin{align}
K_{\sigma}({\textstyle \frac{\nu L}{c}})f_{\text{eq}}\left(\mathbf{x}^{\star},c\,\mathbf{n}\right) & \leq f\left(\mathbf{x}_{1},c\,\mathbf{n}\right)\label{eq:f1Estimate}
\end{align}
with some positive constant $K_{\sigma}\leq1,$ independent of $\mathbf{x}^{\star}$
and $\mathbf{x}_{1}$ for fixed $\sigma$. For $\mathbf{x}_{\sigma}$
close to either $\mathbf{x}_{0}$ or $\mathbf{x}_{1}$ the value of $K_{\sigma}$ is very small
and $K_{\sigma}=\mathscr{O}({\textstyle \frac{\nu L}{c}})$ for $c\rightarrow\infty$
at fixed collision frequency and length. If we assume an asymptotic
decay of $f_{1}(c) := f(\mathbf{x}_1, c\,\mathbf{n})$ in the form 
\begin{align}
f_{1}(c)\leq R_{1}\exp\left(-\frac{c^{2}}{2\mathcal{\theta}_{1}^{(\text{tail})}}\right) & \qquad\text{for all }c > r,
\end{align}
the precise condition of Holway reads 
\begin{align}
K_{\sigma}({\textstyle \frac{\nu L}{c}})\frac{\rho^{\star}/m}{(2\pi\theta^{\star})^{3/2}}\exp\left(-\frac{(c-v^{\star})^{2}}{2\theta^{\star}}\right) & \leq R_{1}\exp\left(-\frac{c^{2}}{2\mathcal{\theta}_{1}^{(\text{tail\ensuremath{)}}}}\right)\qquad\text{for all }c > r,
\end{align}
which implies the bound $\theta^{\star}\leq\theta_{1}^{(\text{tail})}$
with the temperature $\theta^{\star}$ at the position $\mathbf{x}^{\star}$.
This is because no matter how small $K_{\sigma}$ becomes for large
velocities, the exponentials are dominating and the inequality is
only satisfied for all $c$, if the exponential decays are behaving
accordingly. Combining this with the decay restriction (\ref{eq:DecayRestriction})
of the Grad expansion at $\mathbf{x}_{1}$ gives
\begin{align}
\theta^{\star} & <2\theta_{1},\label{eq:tempcondition}
\end{align}
a condition involving the actual temperatures of the gas at two separated
positions. 

\subsection{Holway's Original Conclusions and Weiss' Objection}

In the original presentation of the argument in \cite{Holway1964},
Holway considered the one-dimensional situation of a normal shock
wave. Translated to our Fig.~\ref{fig:setup} he assumed a shock
transition between $\mathbf{x}_{1}$ and $\mathbf{x}{}_{\sigma}$.
He is using the coordinates $x_{1}\,\widehat{=}\,\mathbf{x}_{1}$
and $x_{2}\,\widehat{=}\,\mathbf{x}_{\sigma}$ and considers particles
moving in negative direction such that $x_{1}<x_{2}$. His coordinate
$x_{\infty}>x_{2}>x_{1}$ maybe identified with our position $\mathbf{x}_{0}$.
Holway assumes that the shock is so thin that asymptotic shock conditions
can be found already at $x_{1}$ and $x_{2}$. Hence, the distributions
at these points are given by the Maxwellians 
\begin{align}
\lim_{x\rightarrow x_{1}}f(x,\mathbf{c}) & =f_{\mathrm{eq}}^{(1)}(\mathbf{c}):=\frac{\rho_{1}\vert_{\kappa=0}}{(2\pi\theta_{1})^{3/2}}\exp\left(-\frac{|\mathbf{c}-\mathbf{v}_{1}|^{2}}{2\theta_{1}}\right),\label{eq:RightBC}\\
\lim_{x\rightarrow x_{2}}f(x,\mathbf{c}) & =f_{\mathrm{eq}}^{(2)}(\mathbf{c}):=\frac{\rho_{2}\vert_{\kappa=1}}{(2\pi\theta_{2})^{3/2}}\exp\left(-\frac{|\mathbf{c}-\mathbf{v}_{2}|^{2}}{2\theta_{2}}\right),\label{eq:LeftBC}
\end{align}
where the parameters $\rho_{1,2}$, $\mathbf{v}_{1,2}$ and $\theta_{1,2}$
are defined in (\ref{eq:ShockSetting}). Considering our setup in
Fig.~\ref{fig:setup} this means that we find the equilibrium $f_{\mathrm{eq}}^{(2)}$
essentially in all positions $\mathbf{x}_{\sigma}$, $\mathbf{x}^{\star}$
and $\mathbf{x}_{0}$, and $f_{\mathrm{eq}}^{(1)}$ at position $\mathbf{x}_{1}$.
In particular Holway identified for the temperature $\theta^{\star}=\theta_{2}$
and concluded from (\ref{eq:tempcondition}) that $2\theta_{1}>\theta_{2}$
must hold in the boundary conditions for the applicability of Grad's
moment expansion.

The temperatures before and after the shock are connected by Rankine-Hugoniot
conditions depending on the shock's Mach number. Assuming that $\theta_{2}$
belongs to the downstream or hot part \emph{after the shock} and $\theta_{1}$
to the upstream or cold part \emph{before the shock}, Holway computed
a critical Mach number
\begin{equation}
\mathit{Ma^{(\text{crit})}}<\sqrt{\frac{9}{5}+\frac{4\sqrt{6}}{5}}\approx1.939
\end{equation}
beyond which the temperatures before and after the shock would fail
to satisfy condition (\ref{eq:tempcondition}). Actually, Holway \cite{Holway1964}
solved the equation incorrectly and got the limiting Mach number $1.851$,
as has been pointed out by Weiss in \cite{Weiss1996}. 

With this result Holway conjectured that when Grad's moment expansion
is truncated, the range of Mach numbers in which shock solutions exist
is the interval $(1,M)$, with $M$ being a number less than $1.851$.
Holway's original phrasing was ``When the expansion is truncated
after a few terms, \emph{the region of convergence} may be expected
to be smaller than given by {[}the condition{]}'', where the condition
refers to ``$M\le1.851$'' (as mentioned, that number is the result
of a minor miscalculation). Although Holway did not explicitly explain
what he meant by ``the region of convergence'', he did provide a
table showing ``the ranges of convergence'', and the caption of
the table is ``range of Mach numbers for which continuous shock solutions
exist''. Hence, he connected the convergence limit to the sub-shock
artefacts of shock waves as reported by Grad \cite{Grad1952}.

In \cite{Weiss1995}, Weiss found that the 21-moment theory of extended
thermodynamics \cite{Muller1998} predicts smooth shock structures
for any Mach number less than $1.887$, which is in agreement with
the theory of hyperbolic partial differential equations presented
in \cite{Ruggeri}. However, this is beyond Holway's proposed limit.
This inspired Weiss to revisit Holway's proof. Besides finding the
calculational error, he also proposed another objection on Holway's
derivation. He claimed that Holway's argument does not fix the direction
of the shock wave and continues to derive the statement $\theta_{1}\ge\theta^{\star}$
from the relation (\ref{eq:f1Estimate}). Weiss concludes that this
shows that one should set the boundary condition in the opposite way:
$x_{1}$ should point to the hot fluid \emph{behind the shock}, while
$x_{2}\,\widehat{=}\,x^{\star}$ should point to the cold fluid \emph{before
the shock}. In that case condition (\ref{eq:tempcondition}) is naturally
satisfied, because $\theta_{1}>\theta_{2}$ in such a shock and the
restriction on the Mach number is removed. In particular, Weiss concluded
that Holway's argument does not affect the existence of sub-shock
artefacts in shock solutions of moment equations.

\section{Further Discussion}

\subsection{What Holway and Weiss got right and what wrong}

According to our derivation in Sec.~\ref{subsec:Derivation}, there
is no restriction how to place a normal shock wave along the line
from $\mathbf{x}_{0}$ to $\mathbf{x}_{1}$. Weiss was right to point
out that the shock direction could have been chosen differently from
Holway. However, our multi-dimensional derivation also shows that
the temperature condition (\ref{eq:tempcondition}) actually holds
between the temperatures of any two position in any process.

Furthermore, in our view, one cannot conclude from (\ref{eq:f1Estimate})
that $\theta_{1}\geqslant\theta^{\star}$. Note that the ``constant''
$K_{\sigma}$ appearing in that statement actually depends on $\sigma$,
that is, the scaled distance between $\mathbf{x}_{1}$ and $\mathbf{x}^{\star}$,
with $\sigma=1$ corresponding to $\mathbf{x}^{\star}=\mathbf{x}_{0}$
the furthest away from $\mathbf{x}_{1}$. In fact, in the shock scenario
of Holway we find 
\begin{equation}
\lim_{\mathbf{x}^{\star}\rightarrow\mathbf{x}_{0}}f(\mathbf{x}^{\star},\mathbf{c})=f_{\mathrm{eq}}^{(2)}(\mathbf{c}),\qquad\text{and}\qquad\lim_{\mathbf{x}^{\star}\rightarrow\mathbf{x}_{0}}K_{\sigma}=0,\label{eq:Limit}
\end{equation}
hence, no contradiction to (\ref{eq:f1Estimate}). In this sense,
we support Holway's argument on the convergence of Grad's method,
and in view of Sec.~\ref{subsec:Derivation} we also conclude
that Grad's expansion might not converge in a process in which temperature
ratios of more than a factor of 2 are present between any two points.

However, it remains unclear how this relates to the existence of smooth
shock structure. Despite Holway's conjecture, there is no clear evidence
in his argument showing that a smooth shock structure does not exist
for a Mach number larger than $1.939$, especially for a low-moment
theory. Holway's convergence argument can only explain the limiting
diverging behavior of Grad's theory, while it can be seen from
Fig.~\ref{fig:dis} that sometimes an early truncation of Grad's series can
generate modest approximations of the distribution function. Although
Weiss' work \cite{Weiss1995} still focused only on Mach numbers
less than $1.887$, he has extended the same result to the 35-moment
theory, for which smooth shock structures exists for Mach number
less than $2.2$. The results are reported in \cite[Chapter 12, Section 5]{Muller1998}.
In this sense, we support Weiss' argument that the sub-shock problem
cannot be related to the convergence restriction. Indeed, we tend
to believe that low-moment theories may still provide decent predictions
for moderately rarefied gases, despite the possible divergence of
Grad's expansion in the limit of infinitely many moments.

Note, that the occurence of subshocks in shock solutions of moment
equations is extensively explained, for example, in theoretical terms
in the book \cite{Muller1998} and on the basis of a model problem
in \cite{Torr}. The reason lies in the characteristics of the hyperbolic
waves generated by the equation and their interaction with the relaxational
part of the moment system. The highest characteristic velocity of
the system gives the maximal speed with which infinitesimal disturbances
can propagate. An inflow velocity exceeding this speed can only be
sustained in a steady shock solution by introducing a discontinous
sub-shock, which due to its nonlinearity may move faster than the
characteristic limit.

\subsection{Consequences for Moment Equations}

Nowadays, a lot more numerical experiments have been done for Grad's
method with a large number of moments. To obtain results in strong
non-equilibrium, another issue of Grad's equations is the
loss of hyperbolicity \cite{Muller1998}, that has to be fixed. Such an
issue has been addressed systematically in \cite{Cai2014,Cai2015},
which provides hyperbolic versions of Grad's equations for any number
of moments. However, such hyperbolicity fix does not change Grad's
ansatz, and therefore the convergence issue remains.

Based on the hyperbolic moment equations, several numerical experiments
have been carried out in \cite{Cai2013g,Cai2014d,Cai2018}. All the
three works used a large number of moments in the simulation. However,
after examining all the numerical tests in these three works, we find
that for most cases, the maximum temperature ratio in the numerical
solution is less than $2$. There are only three exceptions, all found
in \cite{Cai2013g}: 
\begin{itemize}
\item Shock tube problem with Knudsen number $0.0251$ computed using 20/84
moments. See \cite[Fig. 2]{Cai2013g}. 
\item Fourier flow with Knudsen number $0.0298$ computed using 56 moments.
See \cite[Fig. 10(a)]{Cai2013g}. 
\item Fourier flow with Knudsen number $0.0658$ computed using 56 moments.
See \cite[Fig. 10(b)]{Cai2013g}. 
\end{itemize}
In all the three cases, the Knudsen number is relatively small, so
that the distribution function is close to the local equilibrium.
Meanwhile, note that the numbers of moments presented in the above
list correspond to the three-dimensional case, meaning that an early
truncation of Grad's series is used equivalent to $M\approx4-6$, 
and by our observation in Sec.\,\ref{sec:failure},
it can still be expected that one can get reasonable approximations
of the distribution functions. Additionally, in all the above cases,
the maximum temperature ratio does not exceed $3$, so that the problem
may still be manageable for a small number of moments. As a summary,
existing numerical experiments support our conclusion that moderate
non-equilibrium flow can still be well captured by Grad's method with
a moderate number of moments.

It is worth mentioning that the situation may change once Grad's equations
are linearized \cite{Torrilhon2015}. In the Grad expansion, the nonlinearity
comes completely from the involvement of velocity and temperature
in the distribution function. Therefore, the linearization of Grad's
equations changes the ansatz of the distribution function by replacing
the local equilibrium by a global equilibrium: 
\begin{equation}
f_{LG}^{(M)}(\mathbf{c})=\sum_{|\alpha|=0}^{M}\lambda_{\alpha}\mathbf{c}^{\alpha}f_{\mathrm{global}}(\mathbf{c}),\label{eq:linear}
\end{equation}
where 
\begin{equation}
f_{\mathrm{global}}(\mathbf{c})=\frac{1}{m(2\pi\theta^{(0)})^{3/2}}\exp\left(-\frac{(c_{i}-v_{i}^{(0)})^{2}}{2\theta^{(0)}}\right),
\end{equation}
with $v_{i}^{(0)}$ and $\theta^{(0)}$ being constants. Similar to
(\ref{eq:DecayRestriction}), the convergence of the above expansion
as $M\rightarrow\infty$ requires that $2\theta^{(0)}>\theta^{(\text{tail})}$.
For linearized Grad's equations, this parameter is manually chosen
by setting the global equilibrium about which the linearization is
performed. Therefore, if $\theta^{(0)}$ is chosen sufficiently large
such that $2\theta^{(0)}>\theta^{(\text{tail})}$ throughout the computational
domain, then the convergence can again be achieved regardless of whether
the original Grad's method converges. Such numerical results have
appeared in a recent work \cite{Hu2019} and \cite[Figure 5.7]{Hu2019} shows
the result of the Fourier flow with Knudsen number $0.1$ and up to
$5456$ moments, where the temperature ratio is larger than $2.5$.
No divergence is observed since the value of $\theta^{(0)}$ is chosen
to be even larger than the highest temperature in the numerical result.
Nevertheless, such a method discards the consideration that the distribution
functions are close to local equilibrium for dense gases, and therefore
may lose efficiency of Grad's method in the near-continuum regime.
In fact, any equilibrium distribution with temperature different from
$\theta^{(0)}$ will require a non-trivial expansion (\ref{eq:linear}).

In his dissertation \cite{HolwayDiss} Holway suggests a non-linear
alternative to Grad's expansion specifically for the computation of
normal shock profiles. He modifies the Mott-Smith-ansatz \cite{MottSmith}
of a bimodal distribution by replacing the Maxwellian with the lower
temperature with a Grad expansion. Due to the superposition with the
hotter Maxwellian a hot tail can be approximated at any point in the
shock profile and convergence is recovered. Unfortunately, this approach
hardly generalizes to other multi-dimensional processes in which the
highest temperature is a-priori not known.

Full non-linear expansions like the maximum-entropy distribution \cite{Dreyer1987,levermore1996}
do not rely on the form (\ref{eq:GradExpansion}). Hence, the arguments
of this paper do not apply. However, convergence of the maximum-entropy
distribution in the limit of many moments remains an open problem
in itself.

\subsection{Numerical Evidence}

Since existing results do not explicitly show the failure of Grad's
method, we are going to support our argument by showing some results
for a one-dimensional problem ($d=1$) with boundary conditions. We
assume that the fluid is located between two parallel diffusive walls
with temperature $\theta_{l}$ and $\theta_{r}$, and we want to solve
for steady state. By approximating the collisions using the BGK operator,
the governing equation can be written as 
\begin{equation} 
c\cdot\partial_{x}f(x,c)=\frac{1}{\mathit{Kn}}[f_{\mathrm{eq}}(x,c)-f(x,c)],\qquad\forall x\in(-1/2,1/2),\quad\forall c\in\mathbb{R},
\end{equation}
with boundary conditions 
\begin{align}
 & f(-1/2,c)=\frac{\rho_{l}/m}{\sqrt{2\pi\theta_{l}}}\exp\left(-\frac{c^{2}}{2\theta_{l}}\right),\qquad\forall c>0,\\
 & f(1/2,c)=\frac{\rho_{r}/m}{\sqrt{2\pi\theta_{r}}}\exp\left(-\frac{c^{2}}{2\theta_{r}}\right),\qquad\forall c<0,
\end{align}
where $\mathit{Kn}$ is the Knudsen number, and in the boundary conditions,
$\rho_{l}$ and $\rho_{r}$ are chosen such that 
\begin{equation}
\int_{\mathbb{R}}cf(-1/2,c)\,dc=\int_{\mathbb{R}}cf(1/2,c)\,dc=0,\label{eq:rho_W}
\end{equation}
which ensures the mass conservation. Furthermore, we assume that the
total mass equals $1$: 
\begin{equation}
m\int_{-1/2}^{1/2}\int_{\mathbb{R}}f(x,c)\,dc\,dx=1.\label{eq:mass}
\end{equation}

For this problem, we can find the exact solution in the limits $\mathit{Kn}\rightarrow+\infty$
and $\mathit{Kn}\rightarrow0$. When $\mathit{Kn}\rightarrow+\infty$,
the BGK equation shows that $f(x,c)$ is independent of $x$. Thus
by the boundary condition, we see that 
\begin{equation}
f(x,c)=\left\{ \begin{array}{ll}
f(-1/2,c), & \text{if }c>0,\\
f(1/2,c), & \text{if }c<0,
\end{array}\right.
\end{equation}
The values of $\rho_{l}$ and $\rho_{r}$ can be solved from (\ref{eq:mass})
and (\ref{eq:rho_W}), and the result is 
\begin{equation}
\rho_{l}=\frac{2\sqrt{\theta_{r}}}{\sqrt{\theta_{l}}+\sqrt{\theta_{r}}},\qquad\rho_{r}=\frac{2\sqrt{\theta_{l}}}{\sqrt{\theta_{l}}+\sqrt{\theta_{r}}}.
\end{equation}
The corresponding temperature is $\theta(x)\equiv\sqrt{\theta_{l}\theta_{r}}$.
Consider the distribution function on the boundaries. We see that
a necessary condition for the convergence of the Grad expansion is
$\sqrt{\theta_{l}\theta_{r}}>\theta_{l}/2$ and $\sqrt{\theta_{l}\theta_{r}}>\theta_{r}/2$.
This shows that if $\theta_{l}>4\theta_{r}$ or $\theta_{r}>4\theta_{l}$,
the convergence of the Grad expansion fails if $\mathit{Kn}$ is sufficiently
large.

\begin{figure}[!t]
\centering \subfloat[$\mathit{Kn} = 0.2$]{   \begin{overpic}[height=.23\textwidth]{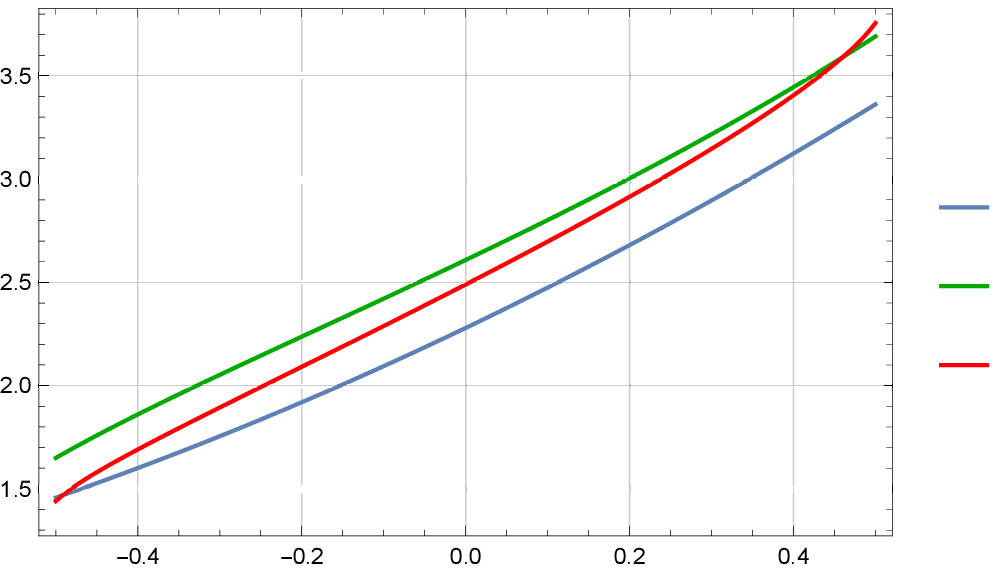}     \put(92.7,17.4){\scalebox{.6}{DVM}}     \put(92.7,32.4){\scalebox{.6}{$M=5$}}     \put(92.7,24.9){\scalebox{.6}{$M=7$}}   \end{overpic}} \hspace{15pt} \subfloat[$\mathit{Kn} = 5$]{   \label{fig:Fourier_Kn5}   \begin{overpic}[height=.23\textwidth]{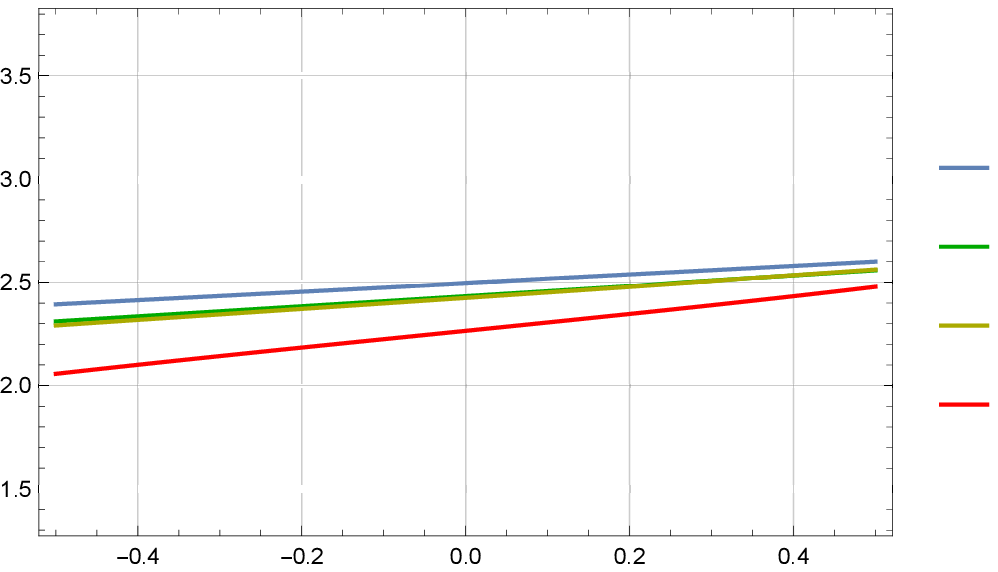}     \put(91.6,13.6){\scalebox{.6}{DVM}}     \put(91.6,21.1){\scalebox{.6}{$M=15$}}     \put(91.6,28.4){\scalebox{.6}{$M=11$}}     \put(91.6,35.7){\scalebox{.6}{$M=7$}}   \end{overpic}} \caption{The temperature profile for discrete velocity model (DVM) and the
moment models for two different cases \textbf{(a)} $\mathit{Kn} = 0.2$ 
and \textbf{(b)}$ \mathit{Kn} = 5$. }
\label{fig:Fourier} 
\end{figure}
To study the limit of the solution as $\mathit{Kn}\rightarrow0$,
we integrate the BGK equation with respect to $c$, which gives us
$v'(x)\equiv0$. By the boundary condition, we see that $v(x)\equiv0$.
Similarly, by multiplying the equation by $c$ and integrating with
respect to $c$, we know that $\rho(x)\theta(x)$ is a constant. Now
we use Chapman-Enskog expansion and assume 
\begin{equation}
f(x,c)=f_{\mathrm{eq}}(x,c)+\mathit{Kn}f^{(1)}(x,c)+\mathit{Kn}^{2}f^{(2)}(x,c)+\cdots.
\end{equation}
By matching the $O(1)$ terms, we get $f^{(1)}(x,c)=c\cdot\partial_{x}f_{\mathrm{eq}}(x,c)$.
We integrate the BGK equation against $c^{2}$ and approximate
$f$ by $f_{\mathrm{eq}}+\mathit{Kn}f^{(1)}$, and the resulting equation,
which is actually the Fourier equation, is 
\begin{equation}
\theta''(x)=0.
\end{equation}
Therefore $\theta(x)$ is approximately a linear function. When $\mathit{Kn}\rightarrow0$,
the distribution function $f(x,c)$ tends to the local Maxwellian.
Therefore by the boundary condition, 
\begin{equation}
\lim_{\mathit{Kn}\rightarrow0}\theta(-1/2)=\theta_{l},\qquad\lim_{\mathit{Kn}\rightarrow0}\theta(1/2)=\theta_{r}.
\end{equation}
Consequently, 
\begin{equation}
\lim_{\mathit{Kn}\rightarrow0}\theta(x)=\left(\frac{1}{2}-x\right)\theta_{l}+\left(\frac{1}{2}+x\right)\theta_{r}.
\end{equation}
This shows that if $\theta_{l}>2\theta_{r}$ or $\theta_{r}>2\theta_{l}$,
we can find a sufficiently small $\mathit{Kn}$ such that the Grad
expansion fails to converge.

To validate our argument that the Grad expansion may still work for
small values of $M$, in the numerical test, we choose the Knudsen
number $\mathit{Kn}=0.2$ and $\mathit{Kn}=5$, and we set the wall
temperatures to be $\theta_{l}=1$ and $\theta_{r}=5$, so that the
Grad expansion will diverge for both small and large Knudsen numbers.
In order to ensure that the number of boundary conditions matches
the number of characteristics pointing insides the domain (which is
necessary for the existence of the solution), we adopt the variation
of Grad's equations with global hyperbolicity \cite{Cai2013}, and
we always choose an odd $M$ since the equations for even $M$ may
have multiple exact solutions. For this one-dimensional problem, the
moment equations are solved with the shooting method with a quasi-Newton
method used for the iteration. The solutions for the temperature field are 
shown in Fig.~\ref{fig:Fourier} together with the result of the discrete velocity
model (DVM) as reference solution.

From the DVM results, we see that for $\mathit{Kn}=0.2$, the temperature
on the right boundary exceeds twice the temperature on the left boundary.
Therefore the Grad expansion is expected to be divergent. When $\mathit{Kn}=5$,
the temperature on the right boundary is less than $\theta_{r}/2$,
so that the Grad expansion also diverges. However, when we solve the
moment equations, the quasi-Newton iteration does not converge for
some $M$. For $\mathit{Kn}=0.2$, the iteration converges only for
$M=5$ and $M=7$, while for $\mathit{Kn}=5$, the iteration diverges
for $M=5,9,13$, but we can find solution for $M=7,11,15$. For large
$M$, we cannot find solution for both Knudsen numbers.

\begin{figure}[!t]
\centering \subfloat[$x = -0.5$]{   \begin{overpic}[height=.23\textwidth]{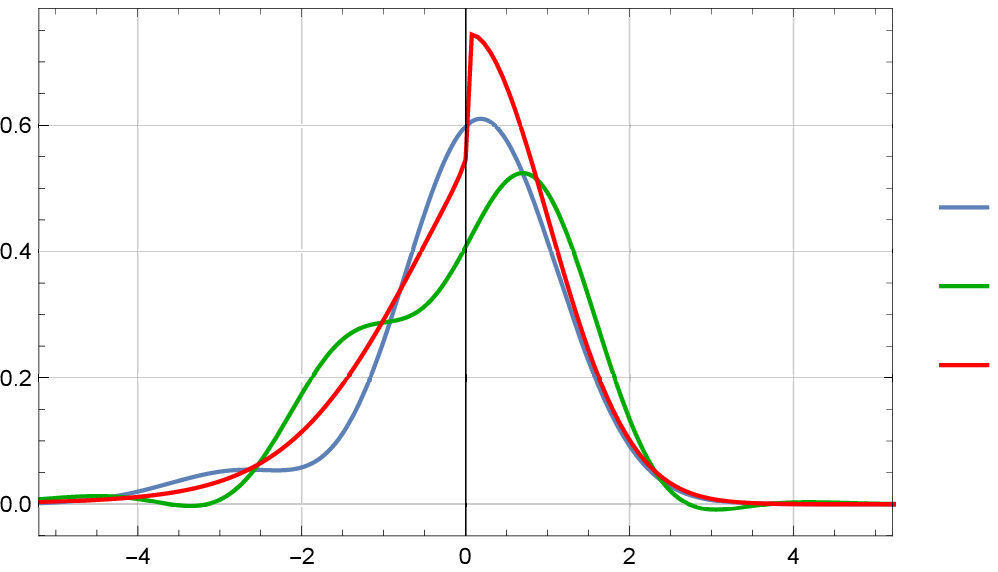}     \put(92.7,17.4){\scalebox{.6}{DVM}}     \put(92.7,32.3){\scalebox{.6}{$M=5$}}     \put(92.7,24.9){\scalebox{.6}{$M=7$}}   \end{overpic}} \hspace{15pt} \subfloat[$x = 0.5$]{   \begin{overpic}[height=.23\textwidth]{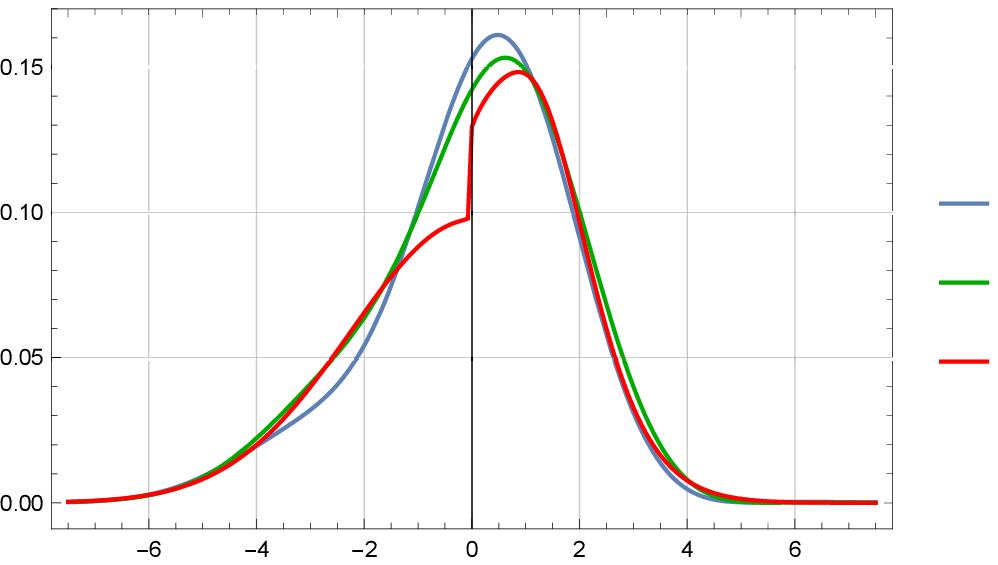}     \put(92.7,17.4){\scalebox{.6}{DVM}}     \put(92.7,32.0){\scalebox{.6}{$M=5$}}     \put(92.7,24.7){\scalebox{.6}{$M=7$}}   \end{overpic}} \caption{Distribution functions at \textbf{(a)} the left boundary, and
\textbf{(b)} the right boundary, for Knudsen number $0.2$}
\label{fig:dis_Kn0p2} 
\end{figure}

From Fig.~\ref{fig:Fourier}, one can also see that all the results are
qualitatively correct. For smaller Knudsen number $\mathit{Kn}=0.2$,
in general, the result of $M=7$ gives better approximation. However,
near the left boundary, where the Grad expansion is expected to be
divergent, $M=7$ gives even larger error than $M=5$. This can be
observed more clearly from the approximation of the distribution functions,
which are plotted in Fig.~\ref{fig:dis_Kn0p2}. It is shown that
on the left boundary, the result of $M=5$ is closer to the DVM result,
whereas on the right boundary, the result of $M=7$ is slightly better.

\begin{figure}[!t]
\centering \subfloat[$x = -0.5$]{   \begin{overpic}[height=.23\textwidth]{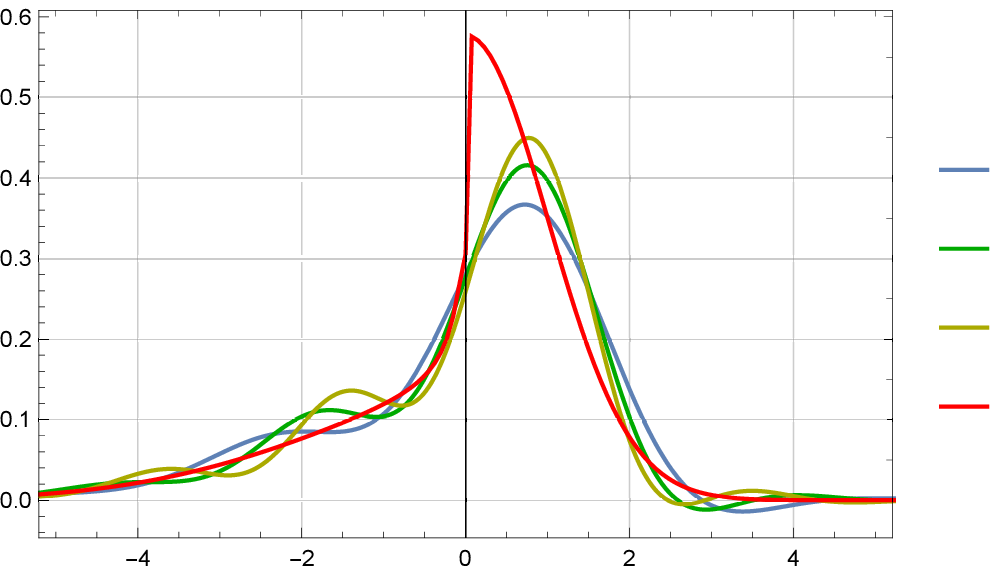}     \put(91.6,13.6){\scalebox{.6}{DVM}}     \put(91.6,21.1){\scalebox{.6}{$M=15$}}     \put(91.6,28.4){\scalebox{.6}{$M=11$}}     \put(91.6,35.7){\scalebox{.6}{$M=7$}}   \end{overpic}} \hspace{15pt} \subfloat[$x = 0.5$]{   \begin{overpic}[height=.23\textwidth]{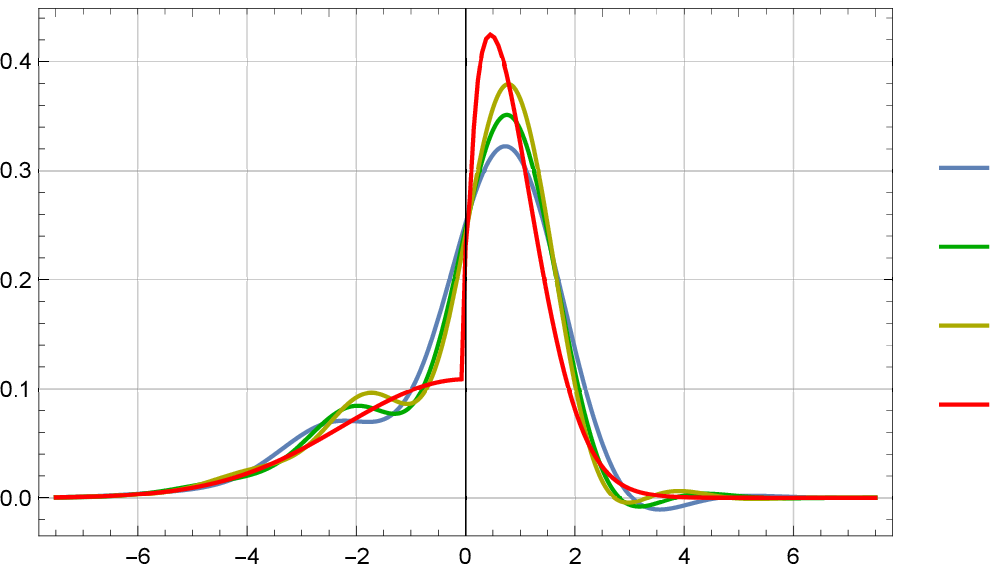}     \put(91.6,13.6){\scalebox{.6}{DVM}}     \put(91.6,21.1){\scalebox{.6}{$M=15$}}     \put(91.6,28.4){\scalebox{.6}{$M=11$}}     \put(91.6,35.7){\scalebox{.6}{$M=7$}}   \end{overpic}} \caption{Distribution functions at 
\textbf{(a)} the left boundary, and
\textbf{(b)} the right boundary, for Knudsen number $5.0$}
\label{fig:dis_Kn5} 
\end{figure}

Similar plots for $\mathit{Kn}=5$ are provided in Fig.~\ref{fig:dis_Kn5}.
In this example, due to the large discontinuity in the exact solution,
it is harder for the moment method to get a good approximation. Nevertheless,
when $M$ is not too large, the moment method still describes the
general profile of the distribution functions.

\section{Conclusion }

We revisited a debate between L.~H.~Holway and W.~Weiss on the convergence
of moment approximations which keeps generating confusion in the community.
In our view Holway was correct to show that there is a limit on the
applicability of Grad's moment expansion and in fact we generalized
his argument to general multi-dimensional steady processes. However,
Holway's attempt to attribute the sub-shock behavior of moment equations
when computing shock profiles to this convergence restriction is wrong.
This misconception has been correctly pointed out by Weiss, whose
smooth shock profile solutions based on moments remain valid. However,
Weiss' more substantial criticism of Holway's argument turned
out to be unfounded in our study.

Roughly speaking, Holway's argument means that whenever there is a
hot spot in a process, fast particles originating from that
spot can be found anywhere in the domain, generating hot distribution
tails that make Grad's expansion diverge. While this bahavior is real
and can be observed in specific computations, its implications for
gas models based on a relatively small number of moments is probably
negligible. We also discussed that even in simulations with many moments,
convergence issues often remain undetected due to stabilizing effects
like dissipation at moderate Knudsen numbers.

\bibliographystyle{siam}
\bibliography{GradConvergence}

\end{document}